# THE EVOLUTION OF VOLUMETRIC VIDEO: A SURVEY OF SMART TRANSCODING AND COMPRESSION APPROACHES

Preetish Kakkar[1] and Hariharan Ragothaman[2]

[1]IEEE Senior Member, USA
preetish.kakkar@gmail.com

[2]athenahealth, USA
hariharanragothaman@gmail.com

## ABSTRACT

*Volumetric video, the capture and display of three-dimensional (3D) imagery, has emerged as a revolutionary technology poised to transform the media landscape, enabling immersive experiences that transcend the limitations of traditional 2D video. One of the key challenges in this domain is the efficient delivery of these high-bandwidth, data-intensive volumetric video streams, which requires innovative transcoding and compression techniques. This research paper explores the state-of-the-art in volumetric video compression and delivery, with a focus on the potential of AI-driven solutions to address the unique challenges posed by this emerging medium.*

## 1. INTRODUCTION

The advent of volumetric video technology has revolutionized the way we capture, process, and consume digital media. Volumetric video, often referred to as "holographic type" media, offers a more immersive and engaging experience by providing 6 degrees of freedom (6DoF), allowing viewers to freely move around and interact with the virtual environment [2]. This technology has significant potential in transforming various multimedia applications, from virtual reality and augmented reality to teleconferencing and virtual tourism.[2] As we stand at the forefront of a new era in visual communication, volumetric video presents unprecedented opportunities for applications ranging from entertainment and education to healthcare and industrial design.

The stage depicted is a state-of-the-art setup for volumetric video capture, featuring a green screen environment crucial for chroma keying techniques. The configuration includes an array of high-intensity lights positioned to provide uniform illumination, essential for reducing shadows and achieving high-quality footage. Multiple cameras, likely arranged around the perimeter, enable 360-degree recording, capturing detailed volumetric data for creating immersive 3D content. One such example of this technology in action is Volumetric Video Capture Studio [Figure 1]. This setup is optimized for applications in virtual reality (VR), augmented reality (AR), and other immersive media experiences, providing a comprehensive solution for capturing dynamic and interactive video content.

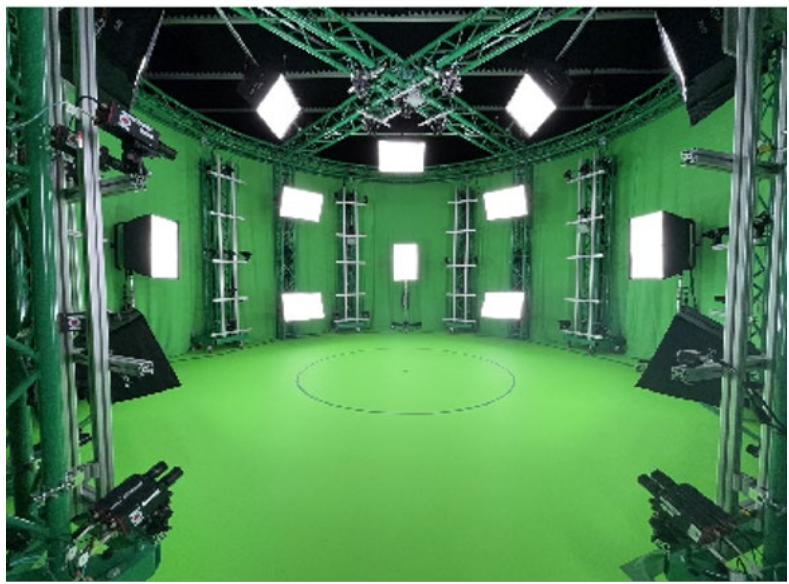

Figure 1 - Volumetric Video Capture Studio

However, the widespread adoption of volumetric video has been hindered by several challenges, including the large file sizes, high computational requirements, and the lack of standardized encoding and delivery mechanisms.

To address these issues, researchers have been exploring smart transcoding techniques that can optimize the quality and efficiency of volumetric video content without compromising the user experience. These techniques involve intelligent compression, adaptive streaming, and efficient encoding algorithms that can adapt to various network conditions, device capabilities, and user preferences. By leveraging advanced transcoding methods, researchers aim to enable the widespread adoption of volumetric video technology by overcoming the challenges of large file sizes, high computational requirements, and the lack of standardized delivery mechanisms.

Recent studies have investigated the potential of point cloud compression and transmission optimization to improve the quality of service for volumetric video streaming. These approaches have shown promising results in reducing the bandwidth requirements and improving the user experience, but they still face limitations in terms of real-time performance and scalability. To further advance the field of volumetric video, researchers are exploring AI-driven solutions that can leverage machine learning and artificial intelligence to optimize the entire pipeline, from capture to display. These AI-driven approaches have the potential to enable more efficient compression, adaptive streaming, and intelligent decision-making, ultimately paving the way for the widespread adoption of volumetric video technology. By harnessing the power of AI, researchers aim to develop innovative techniques that can dynamically adapt to various network conditions, device capabilities, and user preferences, delivering a seamless and immersive volumetric video experience. Through the integration of advanced machine learning algorithms and intelligent data processing, these AI-driven solutions can optimize the encoding, transmission, and rendering of volumetric video content, addressing the challenges of large file sizes, high computational requirements, and the lack of standardized delivery mechanisms. The implementation of AI-driven volumetric video systems can unlock new possibilities in applications such as virtual reality, augmented reality, teleconferencing, and virtual tourism, revolutionizing the way we interact with and consume digital media.

The synergy between volumetric video capture and smart transcoding is paving the way for a future where three-dimensional, interactive content becomes as ubiquitous as traditional video is today. This paper explores the latest advancements in smart transcoding techniques for volumetric video, examining their impact on data compression, transmission efficiency, and overall user experience.

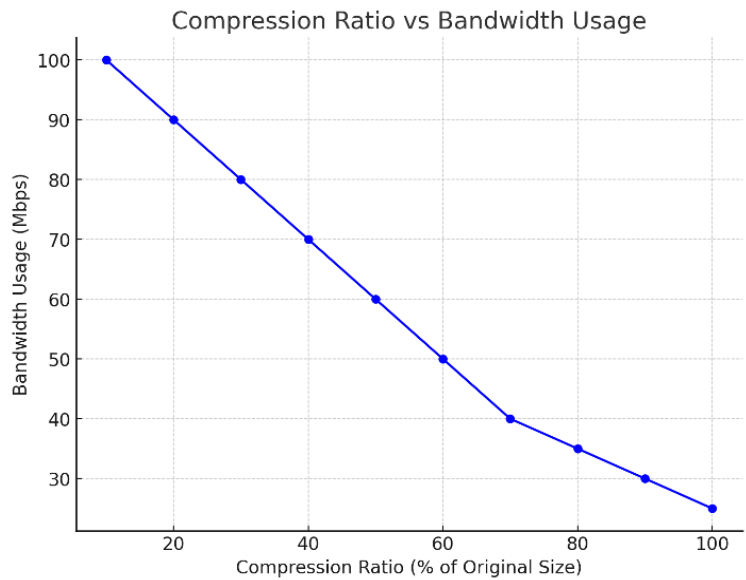

## 2. Emerging Trends and Challenges in Volumetric Video

The rapid advancements in volumetric video technology have led to the emergence of several key trends and challenges that researchers are actively addressing. One of the main challenges in the field of volumetric video is the efficient delivery of high-bandwidth, data-intensive content. Volumetric video data, often represented as point clouds or meshes, can be significantly larger than traditional 2D video, posing significant challenges for real-time streaming and delivery to a wide range of devices. To address this issue, researchers have been exploring advanced compression techniques, such as point cloud compression and dynamic mesh encoding, to reduce the file sizes while preserving the visual quality and interactivity of the volumetric content. Another key challenge is the development of adaptive streaming solutions that can dynamically adjust the quality and resolution of the volumetric video based on the available network bandwidth and device capabilities. Another key challenge is the development of adaptive streaming solutions that can dynamically adjust the quality and resolution of the volumetric video based on the available network bandwidth and device capabilities. These adaptive streaming solutions need to be capable of seamlessly transitioning between different levels of detail and quality, ensuring a smooth and immersive viewing experience for the user, even in variable network conditions.

By implementing intelligent algorithms that can monitor and respond to network fluctuations and device constraints, researchers aim to enable the widespread delivery of volumetric video content across a wide range of platforms and network environments. This involves the development of sophisticated algorithms that can intelligently select the optimal bitrate, resolution, and encoding parameters for the volumetric video stream, balancing the tradeoff between visual quality and bandwidth requirements. Additionally, these adaptive streaming solutions must be capable of dynamically adjusting the content delivery based on real-time feedback from the client devices, ensuring a consistent and high-quality experience regardless of the user's network conditions or device capabilities. By addressing these challenges, researchers can unlock the full potential of volumetric video technology, making it accessible to a wider audience and enabling new immersive experiences in various applications, such as virtual reality, augmented reality, and remote collaboration.

## 3. METHODOLOGY AND METHODS

To address the challenges and unlock the full potential of volumetric video technology, researchers have been exploring a range of methodologies and methods, including:

1. **Adaptive Bitrate Streaming**: One of the key approaches is the development of adaptive bitrate streaming techniques for volumetric video content [4] .These methods involve the use of intelligent algorithms that can dynamically adjust the bitrate, resolution, and encoding parameters of the volumetric video stream based on the available network conditions and device capabilities.
2. **Point Cloud Compression**: Another critical area of research is the development of efficient point cloud compression algorithms that can significantly reduce the file size of volumetric video data without compromising visual quality [4].These compression techniques leverage advanced data structures and encoding methods to optimize the storage and transmission of point cloud data, enabling the delivery of high-quality volumetric content over constrained network environments.
3. **Mesh-based Compression**: In addition to point cloud compression, researchers have also explored mesh-based compression techniques for volumetric video data. These methods exploit the inherent structures and geometric properties of the 3D meshes to achieve efficient compression, enabling the delivery of volumetric content with a smaller footprint.
4. **AI-driven Optimization**: The integration of artificial intelligence and machine learning algorithms has emerged as a promising approach for optimizing the end-to-end delivery of volumetric video [3][1]. These AI-driven solutions can leverage real-time feedback from client devices, network conditions, and user preferences to dynamically adapt the encoding, transmission, and rendering of the volumetric video, ensuring a seamless and high-quality experience.
5. **Standardization and Interoperability**: To enable the widespread adoption of volumetric video technology, researchers are also actively working on the development of standardized formats, protocols, and delivery mechanisms that can ensure interoperability across different platforms and devices [2]

By leveraging these methodologies and methods, researchers aim to overcome the challenges and unlock the full potential of volumetric video technology, enabling immersive and engaging experiences in a wide range of applications.

## 4. SYSTEM DESIGN AND ARCHITECTURE

To enable the efficient delivery of volumetric video content, we propose a system design and architectural approaches that leverage the methodologies. One of the key architectural components is the Volumetric Video Transcoder, which is responsible for converting the high-resolution, data-intensive volumetric video data into more compact and streamable formats [4] [6]. This transcoder can employ advanced compression techniques, such as point cloud compression and mesh-based encoding, to reduce the file size while preserving the visual quality and interactivity of the volumetric content.

The Adaptive Streaming Module is another crucial component that dynamically adjusts the delivery of the volumetric video based on the available network bandwidth and device capabilities. This module utilizes real-time feedback from the client devices to select the optimal bitrate, resolution, and encoding parameters, ensuring a seamless and high-quality viewing experience for the user.

To further optimize the performance and efficiency of the volumetric video delivery, researchers have proposed the integration of AI-driven Optimization Algorithms. These algorithms can leverage machine learning techniques to analyze user preferences, network conditions, and device characteristics, and then make intelligent decisions about the encoding, transmission, and rendering of the volumetric content [4][6] By combining these architectural components, we are working towards the development of end-to-end systems that can effectively deliver volumetric video content to a wide range of devices and network environments, unlocking its potential for various immersive applications[1][4][5].

### 4.1 Volumetric Video Transcoder

The Volumetric Video Transcoder (Figure 2) plays a pivotal role in the system design by converting raw volumetric video data into formats that are more manageable for streaming and storage. Utilizing cutting-edge compression techniques, such as point cloud compression and mesh-based encoding, the transcoder minimizes file sizes without sacrificing the fidelity of the visual content. For instance, [4] demonstrated that point cloud compression could significantly reduce the data size while maintaining high visual quality, making it feasible for real-time applications.

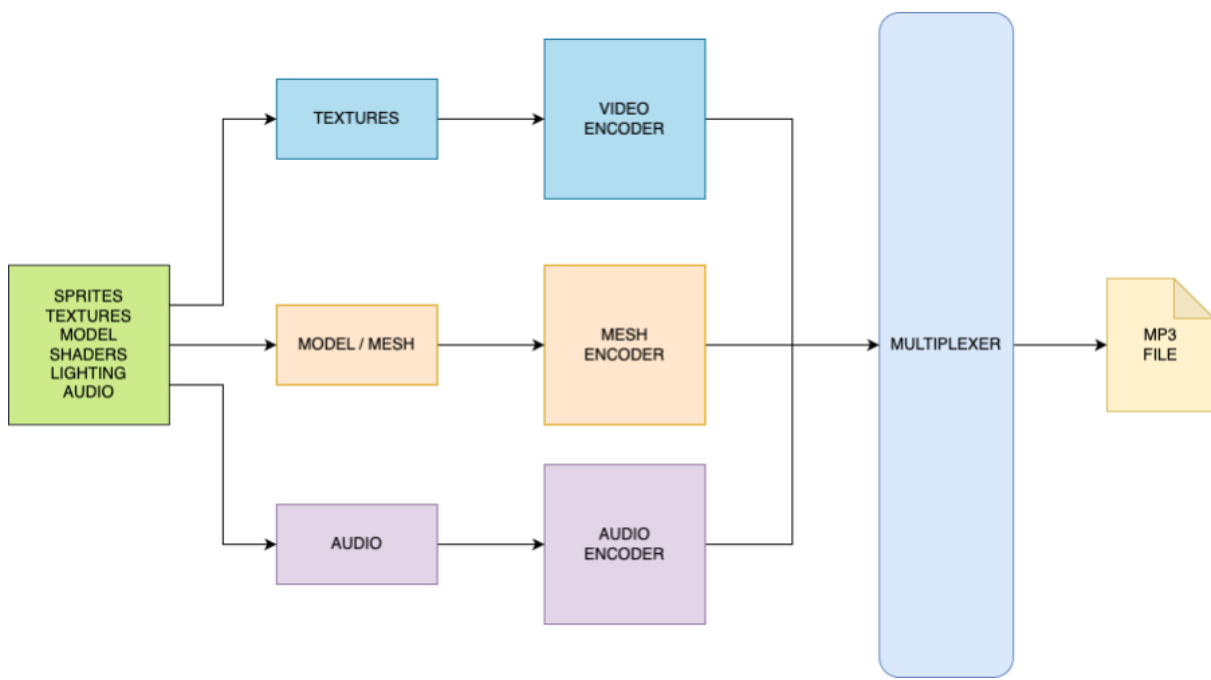

Figure 2 Mesh based Volumetric video process.

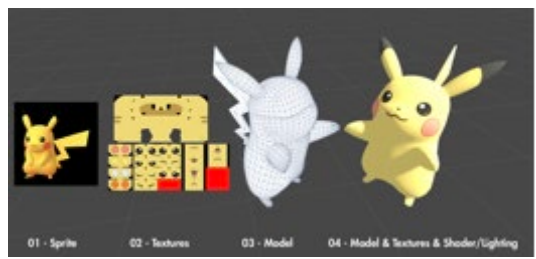

Moreover, the transcoder supports a variety of encoding standards to ensure compatibility with different devices and platforms. By converting volumetric data into widely accepted formats like MP4, the system ensures broad accessibility and ease of integration into existing media ecosystems. This conversion process is depicted in Figure 1, where textures, meshes, and audio are encoded separately and then multiplexed into a single, cohesive output file.

### 4.2 Adaptive Streaming Module

The Adaptive Streaming Module is designed to enhance the user experience by dynamically adjusting the video delivery parameters based on current network conditions and device capabilities. This module employs real-time feedback mechanisms to continually monitor and adapt to changes in bandwidth availability and device performance. Techniques such as adaptive bitrate streaming (ABR) are utilized to ensure that the video quality adjusts seamlessly in response to fluctuating network conditions, thus preventing buffering and other interruptions.

In addition to bitrate adjustments, the Adaptive Streaming Module also optimizes other encoding parameters such as resolution and frame rate. By tailoring these settings to the specific capabilities of the user's device, the module ensures an optimal balance between video quality and performance. For example, high-end devices with powerful processors and high-resolution displays can receive higher quality streams, while lower-end devices are provided with streams that are optimized for their hardware constraints.

### 4.3 AI-Driven Optimization Algorithms

To further enhance the system's efficiency, AI-driven Optimization Algorithms are integrated into the architecture. These algorithms use machine learning techniques to analyze a vast array of data points, including user preferences, historical viewing patterns, network conditions, and device characteristics. By leveraging this data, the system can make intelligent decisions about encoding, transmission, and rendering processes.

For instance, if the system detects that a user frequently experiences network congestion during certain times of the day, it can preemptively adjust the streaming parameters to mitigate the impact of these conditions. Additionally, by understanding user preferences, the system can prioritize certain types of content or visual features, ensuring a more personalized and engaging viewing experience.

### 4.4 Integration and Deployment

By combining these architectural components, we aim to develop a robust and scalable system capable of delivering high-quality volumetric video content across diverse environments. The integration of the Volumetric Video Transcoder, Adaptive Streaming Module, and AI-driven Optimization Algorithms creates a comprehensive solution that addresses the key challenges associated with volumetric video delivery.

The deployment of this system involves setting up the necessary infrastructure to support real-time encoding, streaming, and playback of volumetric content. This includes the establishment of powerful servers for data processing, efficient content delivery networks (CDNs) for distribution, and intuitive client applications for playback. Through rigorous testing and continuous optimization, we strive to ensure that the system meets the demands of both content creators and consumers, paving the way for widespread adoption of volumetric video technology.

## 5. SMART TRANSCODING FRAMEWORKS

### 5.1 EasyVolcap Framework

The EasyVolcap library has emerged as a pivotal tool in volumetric video research, offering a comprehensive framework for neural dynamic scene representation and reconstruction. This open-source project, developed by a team of computer vision and graphics researchers, provides a unified platform for various volumetric video tasks, including capture, reconstruction, and rendering.

EasyVolcap's primary contribution lies in its implementation of state-of-the-art neural rendering techniques. The framework utilizes implicit neural representations to encode complex 3D scenes efficiently. This approach allows for high-quality reconstruction of dynamic scenes from multi-view video inputs, significantly reducing the storage requirements compared to traditional mesh-based representations [4]

One of the key features of EasyVolcap is its support for neural radiance fields (NeRF) and its variants. NeRF-based methods have shown remarkable results in synthesizing novel views of static scenes, and EasyVolcap extends this concept to dynamic scenes. By incorporating temporal coherence and deformation models, the framework enables the reconstruction of moving subjects with high fidelity [4].

The smart transcoding capabilities of EasyVolcap are particularly evident in its ability to compress volumetric video content. By leveraging neural compression techniques, the framework can achieve high compression ratios while maintaining visual quality. This is crucial for streaming applications, where bandwidth constraints often pose significant challenges [4]

### 5.2 DeformStream Framework

DeformStream represents a significant leap forward in volumetric video streaming technology. This innovative approach enhances streaming performance by utilizing deformable mesh-based representations, striking a balance between compression efficiency and rendering quality.

At the core of DeformStream is the concept of representing dynamic 3D scenes as a series of deformable meshes. Unlike traditional static mesh representations, DeformStream allows for efficient encoding of temporal changes in the scene geometry. This is achieved through a combination of keyframe meshes and inter-frame deformation fields.

The smart transcoding process in DeformStream begins with the extraction of keyframe meshes at regular intervals. These keyframes serve as anchor points for the scene geometry. Between keyframes, the system computes and encodes compact deformation fields that describe how the mesh vertices move and deform over time. This approach significantly reduces the amount of data that needs to be transmitted, as only the changes in geometry are encoded rather than full mesh information for each frame.

DeformStream's transcoding pipeline also incorporates adaptive resolution techniques. By analyzing the scene content and viewer perspective, the system can allocate more resources to visually important regions while reducing detail in less critical areas. This content-aware approach further optimizes bandwidth usage without compromising the perceived quality of the volumetric video.

One of the key advantages of DeformStream is its compatibility with existing rendering pipelines. The deformable mesh representation can be efficiently rendered on a wide range of devices, from high-end PCs to mobile devices, making it a versatile solution for volumetric video streaming.

### 5.3 Hierarchical Progressive Coding (HPC) Framework

The Hierarchical Progressive Coding (HPC) framework addresses one of the most significant challenges in volumetric video: the enormous data volume associated with high-quality 3D content. HPC introduces a multi-resolution approach to volumetric video compression and transmission, enabling scalable streaming solutions that can adapt to varying network conditions and device capabilities.

The HPC framework operates on the principle of progressive refinement. It encodes volumetric video content into multiple layers of increasing detail and quality. The base layer contains a low-resolution representation of the scene, which can be quickly transmitted and rendered to provide an initial preview. Subsequent enhancement layers add more geometric detail, texture information, and temporal fidelity.

This hierarchical structure allows for flexible streaming strategies. In bandwidth-constrained scenarios, only the lower layers may be transmitted, ensuring a basic level of content delivery. As network conditions improve or more powerful devices are used, additional layers can be streamed to enhance the visual quality progressively.

The smart transcoding process in HPC involves sophisticated data analysis to determine the optimal distribution of information across layers. Machine learning techniques are employed to identify visually salient features and prioritize their encoding in lower layers. This ensures that even at reduced quality levels, the most important aspects of the scene are preserved.

HPC also incorporates advanced entropy coding techniques to maximize compression efficiency. By exploiting spatial and temporal correlations within and between layers, the framework achieves high compression ratios without introducing significant artifacts. This is particularly important for maintaining the integrity of fine geometric details and surface textures in volumetric video content.

The scalability offered by the HPC framework extends beyond just quality adaptation. It also enables spatial and temporal scalability, allowing for efficient viewport-dependent streaming in immersive applications. By selectively transmitting only the layers relevant to the current user viewpoint, HPC can significantly reduce the required bandwidth while maintaining a high-quality user experience.

In conclusion, these smart transcoding advancements—EasyVolcap, DeformStream, and the HPC framework—represent significant strides in making volumetric video more accessible and practical for a wide range of applications. By addressing key challenges in data compression, transmission efficiency, and adaptive streaming, these technologies are paving the way for the next generation of immersive media experiences.

## 6. APPLICATIONS

### 6.1 Virtual and augmented reality

The advent of smart transcoding techniques has revolutionized the landscape of volumetric video, particularly in its applications within virtual and augmented reality (VR/AR) environments. This section explores how these advancements are reshaping real-time applications in VR and AR, paving the way for more immersive and interactive experiences.

One of the key advantages of smart transcoding in VR and AR applications is its ability to provide perspective-corrected 3D scene visualization to each user. This personalized view ensures that the volumetric content adapts to the user's position and movement, enhancing the sense of presence and immersion within the virtual environment. Such capability is particularly valuable in multi-user scenarios, where each participant can perceive the scene from their unique vantage point, fostering a more natural and intuitive interaction with the virtual content and other users.

### 6.2 Gaming and Interactive Entertainment

The gaming industry stands at the forefront of benefiting from smart transcoding in VR and AR environments. By enabling more immersive and responsive gaming experiences, this technology is transforming how players interact with virtual worlds and each other. Smart transcoding allows for the real-time transmission and rendering of complex 3D models and environments, creating a level of visual fidelity and interactivity previously unattainable in VR and AR gaming.

One notable application of this technology is in the creation of 3D gaming scenarios where users are captured and represented as high-quality 3D models within the game environment. This approach, demonstrated by Doumanoglou et al., utilizes a multi-camera setup with Kinect devices to capture RGB-D frames asynchronously. These frames are then processed on a centralized server to generate colored meshes of the players, complete with skeleton data for accurate motion tracking.

The use of server-based networking schemes for transmitting these 3D representations, coupled with frame-by-frame compression using static mesh encoders, allows for efficient data transfer and rendering. This technological framework enables players to see and interact with each other's realistic 3D avatars in real-time, significantly enhancing the social and competitive aspects of VR gaming.

Moreover, the ability to render highly detailed and dynamic 3D environments opens new possibilities for game design. Developers can now create more expansive and intricate virtual worlds, with real-time environmental changes and physics-based interactions that respond naturally to player actions. This level of responsiveness and detail contributes to a more engaging and immersive gaming experience, potentially leading to longer play sessions and increased player satisfaction.

### 6.3 Video Conferencing and Telepresence

The potential of volumetric video, enhanced by smart transcoding, extends beyond gaming into the realm of remote communication and collaboration through VR and AR platforms. As the demand for more immersive and effective remote interaction tools continues to grow, volumetric video offers a promising solution for enhancing video conferencing and telepresence systems.

Smart transcoding enables the capture, transmission, and rendering of 3D representations of conference participants in real-time. This capability allows for the creation of virtual meeting spaces where attendees can interact with each other's volumetric avatars, providing a sense of presence and spatial awareness that traditional 2D video conferencing lacks. The system evaluated by Zioulis et al. demonstrates the feasibility of this approach, offering perspective-corrected 3D scene visualization for each user in various scenarios, including face-to-face meetings and side-by-side coupled navigation. The implementation of such systems involves a complex pipeline of data capture, processing, and transmission. Multiple Kinect cameras are employed to obtain mesh models of users, with RGB-D frames captured asynchronously and sent to a centralized server for conversion into colored meshes. The inclusion of skeleton data allows for accurate tracking of user movements, further enhancing the realism of the virtual interactions.

This technology has the potential to significantly improve remote collaboration in various professional fields. For instance, in industries such as healthcare, engineering, or design, volumetric video conferencing could allow experts to examine and manipulate 3D models together in a shared virtual space, regardless of their physical locations. This level of interaction could lead to more efficient problem-solving, faster decision-making processes, and improved outcomes in collaborative projects.

Furthermore, the application of volumetric video in telepresence systems could revolutionize remote education and training. By creating immersive virtual classrooms or training environments, educators and trainers can provide more engaging and effective learning experiences. Students or trainees could interact with 3D models, simulations, and each other in ways that closely mimic in-person interactions, potentially improving knowledge retention and skill acquisition. As smart transcoding techniques continue to evolve, we can expect further improvements in the quality, efficiency, and accessibility of volumetric video applications in VR and AR. These advancements will likely lead to more widespread adoption of volumetric telepresence solutions across various industries, fundamentally changing how we approach remote communication and collaboration in the digital age.

## 7. CONCLUSIONS

In conclusion, the proposed system design and architectural approaches represent significant advancements in the field of volumetric video delivery. By leveraging state-of-the-art compression techniques, adaptive streaming methodologies, and AI-driven optimization, we aim to unlock the full potential of volumetric video for a broad range of immersive applications.